\documentstyle[editedvolume,epsfig,numreferences]{crckapb} 

\def\etal{{\it et al.}}
\def\cl{CL 0016+16}
\def\asca{{\it ASCA}}
\def\rosat{{\it ROSAT}}

\begin{opening}
\title{Hot Electrons and Cold Photons: Galaxy \protect\\
Clusters and the Sunyaev-Zel'dovich Effect}\\

\author{J.P.\ HUGHES}
\institute{Rutgers University\\
           Department of Physics and Astronomy\\
           P.O.\ Box 849, Piscataway, NJ 08855-0849 USA}

\end{opening}

\runningtitle{Hot Electrons \& Cold Photons}

\begin{document}


\section{Introduction}
The hot gas in clusters of galaxies emits thermal bremsstrahlung
emission that can be probed directly through measurements in the X-ray
band. Another probe of this gas comes from its effect on the cosmic
microwave background radiation (CMBR): the hot cluster electrons
inverse Compton scatter the CMBR photons and thereby distort the
background radiation from its blackbody spectral form.  Although this,
the Sunyaev-Zel'dovich (SZ) effect, is quite small, heroic efforts
during the 1980's resulted in its detection in three moderately
distant clusters of galaxies: A665, A2218, and \cl.  It is well known
that one of the purposes of conducting such measurements is to
determine the Hubble constant.  The technique has generated
considerable interest because it is independent of all other rungs of
the cosmic distance ladder and is effective over a wide range of
redshifts: $\sim$0.02 to $\sim$1.

\par

In the last few years, the development of sensitive new instruments
for measuring the SZ effect in clusters has sparked a revolution in
the field. Current radio interferometric arrays can now detect and map
the SZ effect in even distant clusters ($z\sim 1$). Another important
development in this field was the launch of the \asca\ satellite with
its broadband X-ray imaging and spectroscopy that allows, for the
first time, accurate determination of gas temperatures in distant
galaxy clusters.  This information is critically important to the
interpretation of the SZ effect, since the determination of $H_0$
depends on the square of the cluster gas temperature.  In the
following I report on the progress that has been made in determining
the cosmic distance scale from the SZ effect and I highlight what has
been learned about galaxy clusters from these investigations.

\section{Current Results}

There are eight galaxy clusters with published measurements of the SZ
effect based on single-dish radiometry, infrared bolometry, or radio
interferometry. Table 1 summarizes the derived $H_0$ values (and 68\%
confidence level errors) from the several clusters, ordered by
increasing redshift and determined under the following assumptions:
\begin{enumerate}
\item spherical symmetry
\item gas density distribution given by $n_e = n_{e0}
\bigl[ 1 + (\theta/\theta_C)^2 \bigr]^{-3\beta/2}$ 
\item isothermal gas distribution
\item unclumped
\item $\Omega_0 = 2 q_0 = 0.2$
\end{enumerate}

\noindent
Relativistic corrections for the CMBR intensity change and the
X-ray bremsstrahlung spectral emissivity, which result in reductions
of order 10\% in the derived $H_0$ values, are both included.

\begin{table}[t]
\begin{center}
\caption{Summary of X-Ray/SZ Effect $H_0$ Measurements}
\begin{tabular}{lccl}
\hline
Cluster & $z$ & $H_0$ ($\rm km\,s^{-1}\,Mpc^{-1}$) & Reference \\
\hline
Coma         & 0.0232 & $64^{+25}_{-21}$ & Herbig et al.~1996 \\
Abell 2256   & 0.0581 & $68^{+21}_{-18}$ & Myers et al.~1997 \\
Abell 478    & 0.0881 & $30^{+17}_{-13}$ & Myers et al.~1997 \\
Abell 2142   & 0.0899 & $46^{+41}_{-28}$ & Myers et al.~1997 \\
Abell 2218   & 0.171  & $59 \pm 23$      & Birkinshaw \& Hughes 1994 \\
Abell 2218   & 0.171  & $35^{+16}_{-15}$ & Jones 1995  \\
Abell 665    & 0.182  & $46 \pm 16$      & Hughes \& Birkinshaw 1998 \\
Abell 2163   & 0.201  & $56^{+39}_{-22}$ & Holzapfel et al.~1997 \\
\cl\         & 0.5455 & $47^{+23}_{-15}$ & Hughes \& Birkinshaw 1997 \\
\hline
\end{tabular}
\end{center}
\vspace{-0.25in}
\end{table}

\par
The average value of these nine measurements weighted by the
individual errors is $H_0 = 48.5 \pm 6.5\,\rm
km\,s^{-1}\,Mpc^{-1}$. However, this value is potentially quite
strongly biased by systematic effects, as I discuss below.

\section{Systematic Uncertainties}

\par
Birkinshaw \& Hughes (1994) and Holzapfel \etal\ (1997) allowed for
large scale radial temperature gradients when analyzing the SZ effect
and X-ray data
for A2218 and A2163, respectively. Both groups found that, for
temperature profiles that fell with radius, the value of $H_0$ derived
under an isothermal assumption would underestimate the true $H_0$
value by 20\%--30\%.

%

\par 

If cluster gas is clumped, then X-ray emissivity will be increased
relative to SZ by a factor greater than unity.  In this case the value
of $H_0$ derived assuming an unclumped gas distribution will be an
upper limit to the true $H_0$ value. Holzapfel \etal\ (1997) used
X-ray spectral fits to constrain the amount of isobaric clumping in
A2163 and found that a reduction in $H_0$ of only $\sim$10\% from the
unclumped case was allowed.

\par

The peculiar motion of clusters relative to the Hubble flow introduces
an additional distortion to the CMBR spectrum usually referred to as
the ``kinematic'' SZ effect.  For a cluster with a peculiar velocity
of $1000\,\rm km\,s^{-1}$ and temperature of $10\,\rm keV$ the
strength of the kinematic SZ effect would be 9\% of the thermal effect
in the Rayleigh-Jeans portion of the CMBR spectrum. Since the SZ
effect intensity enters as a square in the equation determining $H_0$,
the kinematic SZ effect could introduce up to a $\sim$$\pm$20\% correction.
Peculiar velocities are unlikely to be correlated for clusters that
are widely distributed in redshift and position, so this effect would
result in an additional random uncertainty in $H_0$ for any single
cluster.

\par

It is now clear based on \rosat\ observations that many, if not most,
clusters show evidence for complex surface brightness
distributions. In recent work Hughes \& Birkinshaw (1997) analyze
\cl, a distant cluster that displays strong ellipticity (see
the left panel of Fig.~1). They fit elliptical isothermal-$\beta$
models to the X-ray image and deproject under the assumption that the
three-dimensional structure of the cluster is axisymmetric, either
prolate or oblate. If the symmetry axis is assumed to lie in the plane
of the sky, then the different assumptions about the shape of the gas
distribution yield values for $H_0$ that differ by 17\%, which is the
percentage difference between the major and minor axis lengths of the
cluster.  As the symmetry axis of the ellipsoid is allowed to vary
toward the line-of-sight, then the intrinsic cluster ellipticity
(defined as the ratio of major to minor axis lengths) grows
increasingly larger, as does the uncertainty on $H_0$, which is shown
graphically in the right panel of Fig.~1.  In order to bound the
uncertainty, Hughes \& Birkinshaw argue that it is unlikely for a
cluster to have an intrinsic ellipticity greater than about 1.5, based
on observations of other galaxy clusters. 
This limit results in a uncertainty in $H_0$ of $\sim$$\pm20\%$ for
\cl\ from morphology alone.

\par

To ensure that the effects of unknown geometry and arbitrary
inclination are uncorrelated from cluster to cluster, it is essential
that the cluster sample for determining $H_0$ be selected properly.
For example, as pointed out by Birkinshaw, Hughes, \& Arnaud (1991),
it is important that clusters {\it not} be selected based on the
strength of their SZ effect signal or central X-ray surface brightness,
since this would result naturally in a bias toward prolate clusters
with their long axes aligned to the line-of-sight. Recent cluster
samples for $H_0$ determination have been selected based on the
strength of their integrated X-ray flux from surveys by the {\it
Einstein Observatory} or \rosat. These samples should be relatively
unbiased.

\begin{figure}
\begin{center}
\hbox{
 \epsfig{figure=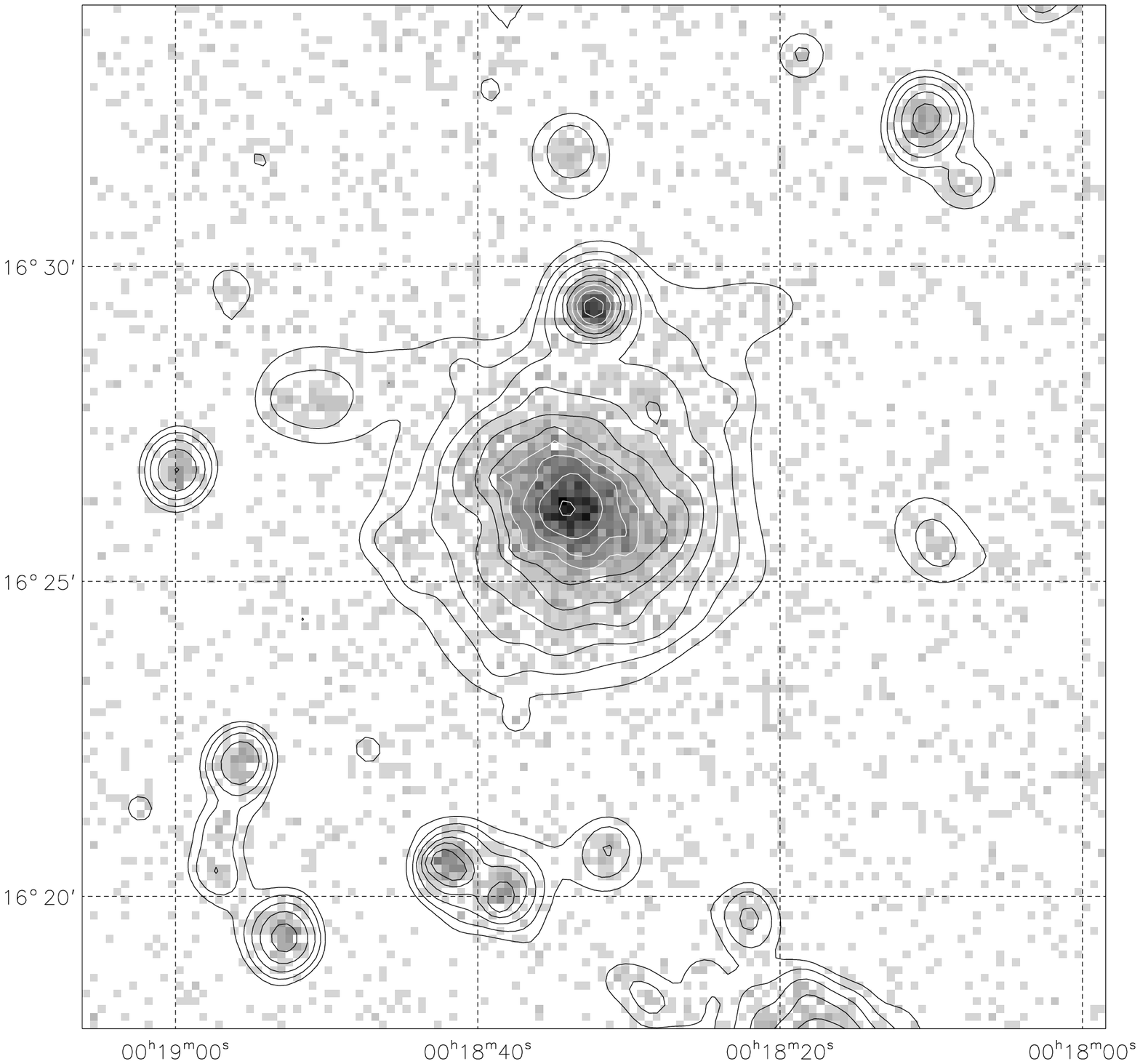,width=2.25in}
 \epsfig{figure=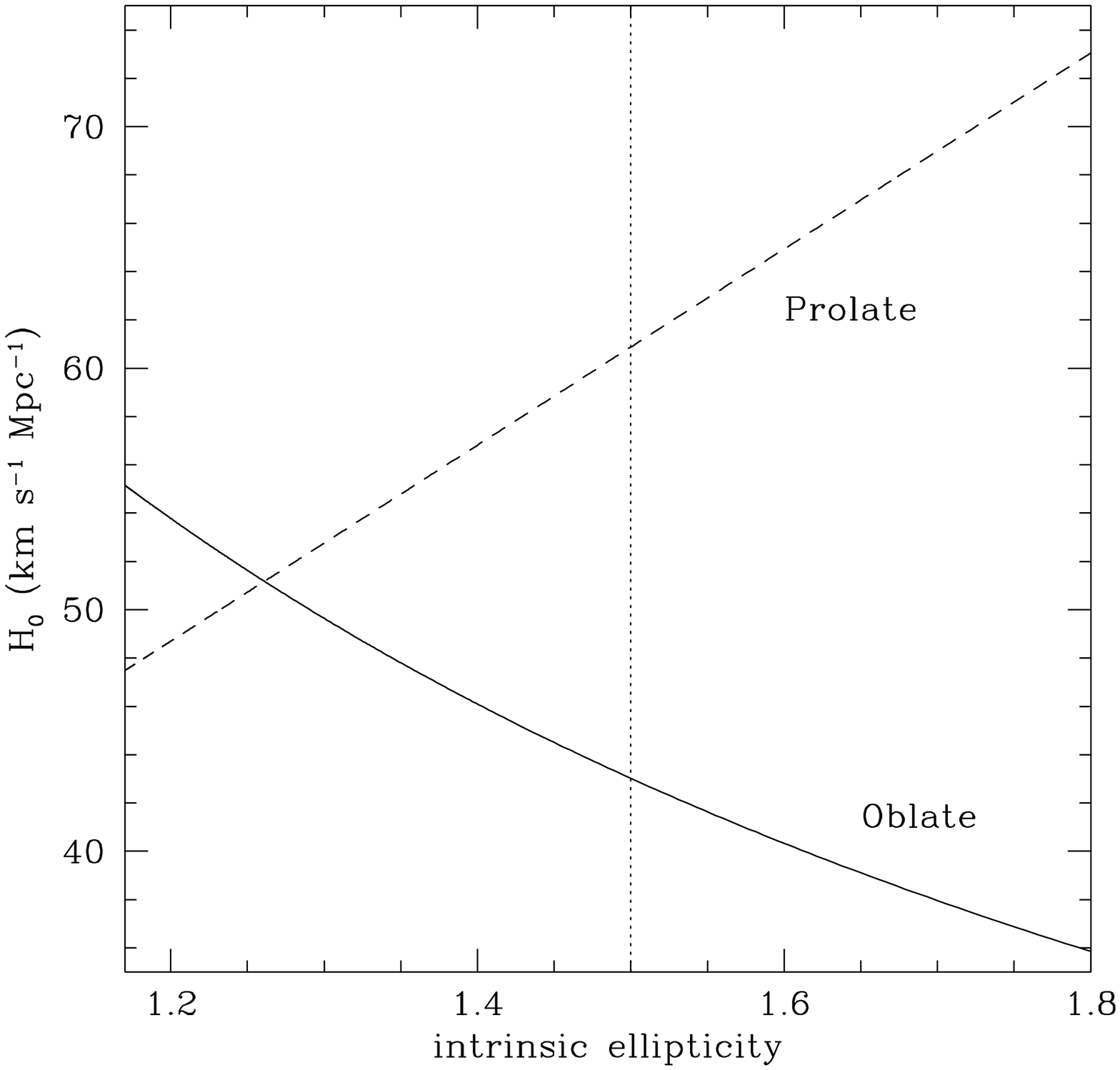,width=2.5in,bbllx=0pt,bblly=175pt,bburx=575pt,bbury=688pt}
}
\end{center}
\vspace{-0.25in}
\caption{(Left panel) {\it ROSAT} PSPC X-ray map of CL 0016+16. (Right
panel) Variation of the derived value of the Hubble constant,
$H_0$, with the intrinsic ellipticity of \cl\ for oblate and prolate
geometries.}
\end{figure}

\section{A Value for the Hubble Constant}

\par

When known systematic uncertainties are included, the best estimate of
the Hubble constant becomes

$$ H_0 = 44 - 64 \,\,\rm km\, s^{-1}\, Mpc^{-1} \,\, \pm 17\%,$$

\noindent
where the range accounts for biases from temperature gradients (+30\%)
and clumped gas ($-$10\%). The quoted 
random error includes observational errors combined in quadrature with
the random systematic errors from peculiar velocities ($\pm$7\%) and
geometry/inclination ($\pm$7\%), which have been reduced from the values
given in \S3 by $1/\sqrt{N}$ where $N$ is the number of clusters. Future
observational efforts should be directed toward measuring the large
scale temperature gradients in galaxy clusters since this is the
single largest uncertainty in the determination of $H_0$ from the SZ
effect.

\def\apj{{\it Ap.J.}}

\par\noindent
{\small This research was partially supported by NASA LTSA Grant
NAG5-3432.}

\end{document}